\documentclass[sigconf,nonacm]{acmart}

\usepackage{subcaption} %
\usepackage{hyperref} %
\usepackage{algorithmic}
\usepackage{graphicx}
\usepackage{textcomp}
\usepackage{xcolor}
\usepackage{framed}
\usepackage{subcaption}
\usepackage{xspace}
\usepackage{cleveref}
\usepackage[frozencache=true,cachedir=_minted-main]{minted}
\usepackage{listings}
\lstdefinestyle{commentcolor}{
  language=Python,
  basicstyle=\ttfamily\fontsize{6pt}{8.5pt}\selectfont,        %
  numbers=left,
  numberstyle=\tiny,
  stepnumber=1,
  breaklines=true,
  breakatwhitespace=true,
  frame=single,
  showstringspaces=false,
  xleftmargin=1.5em,
  keywordstyle={},                         %
  stringstyle={},                          %
  commentstyle=\color{gray}\itshape        %
}

\definecolor{LightGray}{gray}{0.9}

\newenvironment{barquote}{%
  \MakeFramed{\advance\hsize-\width \FrameRestore}%
}{%
  \endMakeFramed%
}

\newcommand{\nbRandomSelectedRepo}{10,000\xspace}
\newcommand{\totalSelectedRepo}{20,000\xspace}

\newcommand{\multistage}{multistage\xspace}

\newcommand{\sizeInputDataset}{27 133 980\xspace}
\newcommand{\DataTwoOneTotal}{35\xspace}

\newcommand{\DataTwoOneYESCount}{31\xspace}

\newcommand{\DataOneOneTotal}{65\xspace}

\newcommand{\DataOneOneYESCount}{35\xspace}
\newcommand{\DataOneOneYESPercent}{53.8\%\xspace}

\newcommand{\DataOneOneNOPercent}{46.2\%\xspace}

\newcommand{\DataOneTwoTotal}{35\xspace}

\newcommand{\DataOneTwoGithubPercent}{65.7\%\xspace}

\newcommand{\DataOneTwoLiteratureDatasetPercent}{40.0\%\xspace}

\newcommand{\DDatasourcesBoxplotMean}{1.46\xspace}
\newcommand{\DDatasourcesBoxplotMedian}{1.00\xspace}

\newcommand{\DataFourOneTotal}{35\xspace}

\newcommand{\DataFourOneYESCount}{24\xspace}

\newcommand{\DataFourOneNOCount}{11\xspace}

\newcommand{\DataFourTwoTotal}{11\xspace}

\newcommand{\DataFourTwoAmbiguityinconstraintdefinitionPercent}{54.5\%\xspace}

\newcommand{\DataFourTwoAmbiguityinMetadataDefinitionPercent}{45.5\%\xspace}

\newcommand{\DataOneThreeTotal}{35\xspace}

\newcommand{\DataOneThreeYESPercent}{65.7\%\xspace}

\newcommand{\DataOneThreeNOPercent}{34.3\%\xspace}

 \newcommand{\swhurl}[1]{https://archive.softwareheritage.org/#1}
\newcommand{\swhref}[2]{\href{\swhurl{#1}}{#2}}

\author[Lefeuvre]{Romain Lefeuvre}
 \affiliation{
 \institution{University of Rennes, Inria, CNRS, IRISA}
 \city{Rennes}
 \country{France}}
 \email{romain.lefeuvre@irisa.fr}
 
 \author[Le Goasteller]{Maïwenn Le Goasteller}
 \affiliation{
 \institution{University of Rennes, Inria, CNRS, IRISA}
 \country{France}
   \city{Rennes}}
 \email{maiwenn.le-goasteller@irisa.fr}
 
 \author[Galasso]{Jessie Galasso}
 \affiliation{
 \institution{McGill University}
 \country{Canada}
 \city{Montréal}}
 \email{jessie.galasso-carbonnel@mcgill.ca}

 \author[Combemale]{Benoit Combemale}
 \affiliation{
 \institution{Inria, University of Rennes, CNRS, IRISA}
 \country{France}
   \city{Rennes}
}
\email{benoit.combemale@inria.fr}

 \author[Perez]{Quentin Perez}
 \affiliation{
 \institution{INSA Rennes, University of Rennes, Inria, CNRS, IRISA}
 \country{France}
   \city{Rennes}
}
\email{quentin.perez@irisa.fr}

\author[Sahraoui]{Houari Sahraoui}
 \affiliation{
 \institution{DIRO, Université de Montréal}
 \country{Canada}
   \city{Montréal }}
 \email{sahraouh@iro.umontreal.ca}

\copyrightyear{2026}
\setcopyright{cc}

\ccsdesc[500]{Software and its engineering~Domain specific languages}

\begin{document}

\title{Modeling Sampling Workflows for Code Repositories}

\begin{abstract}

 Empirical software engineering research often depends on datasets of code repository artifacts, where sampling strategies are employed to enable large-scale analyses. The design and evaluation of these strategies are critical, as they directly influence the generalizability of research findings. However, sampling remains an underestimated aspect in software engineering research: we identify two main challenges related to (1) the design and representativeness of sampling approaches, and (2) the ability to reason about the implications of sampling decisions on generalizability. To address these challenges, we propose a Domain-Specific Language (DSL) to explicitly describe complex sampling strategies through composable sampling operators. This formalism supports both the specification and the reasoning about the generalizability of results based on the applied sampling strategies. We implement the DSL as a Python-based fluent API, and demonstrate how it facilitates representativeness reasoning using statistical indicators extracted from sampling workflows. We validate our approach through a case study of MSR papers involving code repository sampling. Our results show that the DSL can model the sampling strategies reported in recent literature. 

\end{abstract}
\maketitle

\keywords{Sampling workflow, DSL, Sample representativeness}
\section{Introduction}

Empirical software engineering and machine learning for software engineering (SE) research rely heavily on datasets composed of SE artifacts. 
The proliferation of open-source projects and forge platforms has significantly increased access to a large amount of SE-related data. 
This brings forth many challenges regarding the proper utilization of this data for research purposes, including ensuring the quality and reliability of both the data sources and the data, conducting large-scale dataset mining, or handling these datasets effectively for research studies.
In this paper, we focus on the particular challenges related to defining and assessing sampling methodologies applied to datasets of software repositories. 

Empirical research often relies on \textit{sampling} to avoid conducting experiments or analyses on the entire set of elements of interest (called the \textit{population}). 
Sampling is the process of selecting a subset of elements of interest (a \textit{sample}) from a portion of the population that is accessible  (called the \textit{sampling frame}) by following a defined strategy that may include, for instance, selecting elements randomly or filtering elements according to specific characteristics.

However, most sampling strategies used in SE datasets do not align with this three-tier framework, notably due to the complexity and diversity of SE datasets and their sources~\cite{vidoni_systematic_2022}.
In practice, obtaining a sample often involves multiple filtering steps and the creation of intermediate datasets, leading to complex, potentially multistaged sampling processes where the distinctions between population, sampling frame, and sample are not always clear~\cite{baltes_sampling_2022}.
For instance, consider a fictitious study where researchers aim to analyze the commit history of active open-source software (OSS) projects. 
They start by filtering an existing public repository archive (e.g., Software Heritage~\cite{pietri_swh}) to include only projects with commits after January 1, 2023. 
Because of the underrepresentation of large OSS projects, they then divide the filtered set into two groups (projects with fewer than 5 contributors and those with more) and randomly sample \nbRandomSelectedRepo projects from each group.
In this example, the final sample of \totalSelectedRepo projects results from several steps, and both the public archive and the filtered set of active repositories could be considered valid sampling frames.

The contributions presented in this paper aim to support researchers in addressing common challenges related to implementing and assessing multistaged sampling strategies in SE datasets.
More specifically, this paper targets sampling strategies selecting sets of software repositories, usually defined as the first step of the process of Mining Software Repositories~\cite{vidoni_systematic_2022}. 
After reviewing definitions of sampling approaches and representativeness arguments, we discuss known challenges associated with sampling in SE research in ~\Cref{background_challenges}. 
These challenges are classified into two main categories: properly defining multistaged sampling strategies  and reasoning about the representativeness of the resulting sample.
To address these challenges, we propose an approach enabling the explicit description of multistaged sampling strategies from a given dataset using a dedicated formalism.
To achieve that, we introduce a domain-specific language (DSL) designed to guide users in constructing sampling strategies as workflows, combining basic sampling operators and explicitly describing constraints through the different stages.
The constructs of this DSL are presented in \Cref{DSL}, along with examples of its use.  
Making explicit the description of sampling strategies provides an opportunity to support reasoning about representativeness through the automated computation of indicators, which can help practitioners to argue about the suitability of the sample size, or the similarity of the characteristic distributions across the different stages of the workflow.
We discuss how to leverage our DSL for representativeness reasoning in \Cref{reasoning}.
We evaluate the applicability and expressiveness of our DSL in \Cref{sec:casestudy} with a case study exploring four research questions on a sample of 65 papers published in MSR. Our results show that \DataOneOneYESPercent (\DataOneOneYESCount papers) perform sampling on code repositories, confirming that the majority of analyzed sampling methodologies involve selecting code repositories. The case study also demonstrates that our DSL can model the sampling strategies reported in recent literature, validating its concept. 
Finally, \Cref{related_work} and \Cref{conclusion} present the related work and the conclusion, respectively.

\section{Background and Motivations \label{background_challenges}}

\subsection{Sampling Strategies \label{techniques}}

Various sampling strategies have been proposed in software engineering research \cite{cochran1977sampling,baltes_sampling_2022}. 
Some sampling strategies are based on probability sampling, such as \textit{simple random sampling} in which each item of the sampling frame has an equal probability of being selected. 
One of its common variations is \textit{systematic random sampling}, where items are selected at regular intervals, starting from a randomly chosen point, to ensure an even distribution throughout the sample. 
Other strategies do not involve probabilistic mechanisms, such as \textit{convenience sampling,} where the sample items are selected based on their availability or ease of access. 
Another common strategy is \textit{purposive sampling}, which refers to a selection based on the researcher's expertise to identify relevant items to constitute the sample. 
Some strategies require several steps to define a sample: we refer to them as \textit{\multistage sampling strategies}.
For instance, a simple random sampling is not adapted to the study of underrepresented items that have a particular property of interest, since random sampling will result in a low presence of those items. 
\textit{Stratified sampling} consists in extracting items from different population strata, which could be defined on an attribute or item property, then selecting an item from each stratum, either with a random selection (\textit{stratified random sampling}) or purposively (\textit{quota sampling}). In the rest of the paper, we will refer to stratum/strata as subgroups of a sampling frame that are homogeneous for a given characteristic. 
\textit{Cluster sampling} is also a \multistage strategy that involves selecting entire groups, or clusters, within the population. 
The different items in the same cluster are heterogeneous.

Empirical software engineering research that relies on datasets of software artifacts largely uses open-source software repository forges as primary data sources~\cite{vidoni_systematic_2022}.
These repositories typically contain various types of artifacts (e.g., source code, documentation, bug reports)~\cite{vidoni_systematic_2022} interconnected in complex ways~\cite{ma2018automatic,pfeiffer2020constitutes}.
Also, the structure and data models of these artifacts vary depending on the forge that hosts the repository~\cite{lefeuvre_fingerpring}.
Due to this complexity, sampling software artifacts from such data sources typically involves multiple stages of refinement and selection; this results in the need to apply \textit{\multistage sampling strategies}~\cite{baltes_sampling_2022}, which may combine various sampling and clustering techniques.
In our running example, the sampling strategy is \multistage: it combines a \textit{purposive sampling} (the researchers decided that repositories with at least one commit in the last 6 months were considered active) with a \textit{stratified random sampling} on the number of collaborators.

When several stages are involved, the input dataset undergoes several transformation steps, resulting in multiple ``intermediate datasets''.
For instance, in our running example, we obtain a first subset of the input dataset after filtering by the date of the last commit, then split into two more subsets based on the number of contributors. 
These two last subsets are then randomly sampled to form two new subsets, before being merged into the final sample.
In such situations, the straightforward application of the three-tier sampling framework discussed earlier becomes difficult:
indeed, it is unclear which of the intermediate datasets should be considered the population, sampling frame, or sample, since several of them may fulfill these roles at different stages of the  process~\cite{baltes_sampling_2022}.

\textit{This calls for a more flexible approach to representing \multistage sampling strategies, one that can capture their 
variability and complexity.} 
In what follows, we posit that sampling strategies involving several stages are better conceptualized as \textit{workflows}, where each node represents a collection of items (intermediate datasets) and each arc defines a transformation operation (e.g., filtering, sampling) between different states of the collection. 
\Cref{fig:flow} (discussed in more detail in ~\Cref{reasoning}) shows a workflow modeling the \multistage sampling strategy of our running example.

\subsection{Generalization Reasoning}

\begin{figure*}[ht]
    \centering
    \includegraphics[width=0.80\linewidth]{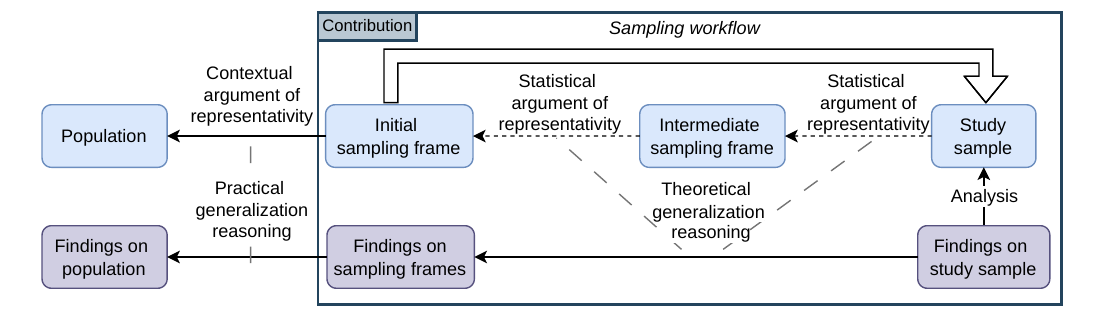}
    \caption{Different levels of generalization reasoning with different type for representativeness argument}
    \label{fig:generalization}
    \Description{Different levels of generalization reasoning with different type for representativeness argument}
\end{figure*}

The aim of sampling is to enable the generalization of the results obtained from the sample to the population as a whole.
Generalizing from a sample to the population can be understood as involving two distinct levels of generalization reasoning.
Figure~\ref{fig:generalization} illustrates these two levels (second row) and their relationships with the concepts of \textit{population}, \textit{sampling frames}, and \textit{samples} (first row), which have been refined to better suit \multistage sampling contexts as follows:

The \textbf{population} refers to the ideal set of items of interest for a study, which may not be fully enumerable or directly accessible. The population is the ultimate target of the study that we ideally aim to generalize to.
For instance:  \textit{all software repositories.}

A \textbf{sampling frame} is a subset of the population that is practically accessible for a study.
For instance, a sampling frame of the previous population could be: \textit{software repositories archived by Software Heritage.}
In \multistage sampling workflows, we perform a series of selections, where each stage produces a subset of items. These subsets can be viewed both as samples of the previous sampling frame and as sampling frames for the next stage.
We thus distinguish between the \textbf{initial sampling frame}, which is the largest accessible frame in a given study, and \textbf{intermediate sampling frames}, which are the multiple subsets generated throughout the workflow and are strictly included in the frame of the previous stage.
Intermediate sampling frames could be \textit{the active software repositories on Software Heritage} or \textit{the active software repositories on Software Heritage and having less than five contributors}.

A \textbf{sample} is a subset of a sampling frame. As noted above, intermediate sampling frames can be viewed as samples of their respective preceding frames.

The final sample obtained at the end of a sampling workflow is considered as the \textbf{study sample}, i.e., the actual dataset used for analysis in a given study.
In our running example, the study sample is the union of \textit{\nbRandomSelectedRepo active repositories with fewer than five contributors} and \textit{\nbRandomSelectedRepo active repositories with five or more contributors}.
\newline

The second row of~\Cref{fig:generalization} illustrates the two levels of generalization we encounter when generalizing findings from a sample to a given population.

First, there is a \textbf{theoretical generalization reasoning} between \textit{samples} and \textit{sampling frames}, which may be supported by statistical representativeness arguments. These statistical arguments can be computed by analyzing the sampling workflow.
While statistical significance can be assessed automatically, the generalization of findings from a sample to a sampling frame requires a manual evaluation of practical significance (i.e., effect size). Even if statistical representativeness is ensured, the effect size must still be assessed manually. A low effect size in the sample may lead a researcher to decide not to generalize the findings, whereas a large effect size may support the decision to generalize.
Note that representativeness arguments in a \multistage sampling context can be computed between multiple pairs of (\textit{sample}, \textit{sampling frames}), due to the existence of multiple intermediate sampling frames. In the original three-tier framework, this reasoning takes place only between the study sample and the initial sampling frame.

Second, there is a \textbf{practical generalization reasoning} between the \textit{initial sampling frame} and the \textit{population}. In many cases, not all elements of the population are accessible. For example, private and commercial repositories may not be analyzable. Therefore, this second type of reasoning should rely on contextual representativeness arguments provided by the study designer.

In what follows, 
references to arguments for representativeness and generalizability focus on theoretical generalization reasoning within the context of the sampling workflow.

\textit{Representativeness discussion : \label{rep_samp}}
Representative sampling is a well-documented concern across scientific fields~\cite{kruskal1979representativeII} (including SE~\cite{baltes_sampling_2022}) and is defined as the extent to which \textit{a sample's properties of interest resemble those of the sampling frame}. 
Ensuring representativeness is crucial for studies aiming for generalization of their results as well as any research using sampling mechanisms. 
Representativeness cannot be ensured solely by the sampling technique; it is a mutual property of a sample and its sampling frame, indicating how closely the sample reflects specific characteristics of the sampling frame for a given analysis. 
Thus, representativeness is dimension-specific (being representative for one criterion does not imply it is for others) and depends on how the sample was selected.
Although representativeness cannot be evaluated with certainty, different arguments for representativeness \cite{kruskal1979representativeII} can support its assessment. 

\textbf{Large and random sample: \label{random}}
The first argument for the representativeness of a sample concerns the use of a probabilistic technique to select the sample from an unbiased sampling frame. The idea behind this is to ensure the absence of selective forces. Such an approach enables probabilistic reasoning supported by sampling theory and statistical inference (e.g., confidence intervals).\cite{kruskal1979representativeIII}

\textbf{Breadth of a sample:}
A broad and heterogeneous sample supports the idea that a subpopulation is unlikely to be excluded by the sampling. The breadth could be evaluated by computing the coverage of distinct values or bins in the sample (ratio between number of classes in the sample over the total number of classes). This coverage could also be evaluated in multiple dimensions \cite{Nagappan_2013_diversity}.

\textbf{Similar distribution:}
Another argument for assessing sample representativeness is to compare the distribution of metadata of interest between a sample and its sampling frame. Statistical methods, such as the \textit{Kolmogorov-Smirnov}\cite{massey1951kolmogorov} or \textit{Chi-square Goodness-of-Fit} tests\cite{Pearson1992}, can be used for this purpose. This is particularly relevant in machine learning, where difference in distributions can negatively impact model training and performance~\cite{10.1145/3503509}.

\textbf{Typical or ideal sample:}
If the previous arguments cannot be applied, coarse-level statistics can be used to support a high-level comparison between the sample and the sampling frame. Metrics such as the mean, median, standard deviation, or mode of the sample can be compared to those of the sampling frame.

Since there is no definitive guarantee for representativeness, it is crucial to evaluate sample quality through thoughtful discussions and transparent justifications for representativeness claims~\cite{kruskal1979representativeI}. 
Effective arguments should include quantitative support: for example, a “large and random sample” claim must be backed by sample size data and statistical test, while arguments about “similar distribution” require a comparison of key property distributions between the sample and sampling frame. 
In the case of \multistage sampling strategies, these arguments have to be applied beyond the usual mapping from population to sampling frame to sample. 
We argue that \textit{it necessitates a shift in how we assess sample representativeness and the generalizability of results, now requiring us to consider how each transformation of the dataset influences the conclusions drawn from the final sample.}
In other words, when several stages and intermediate datasets are involved, arguments for representativeness must be discussed and propagated across the different transformation operations in the sampling workflow, depending on the type and parameters of the corresponding operators.

\subsection{Objectives and Contributions}\label{rep_challenge}
Research on how software engineering studies construct datasets from software repositories suggests that sampling strategies are often not clearly described, and discussions for representativeness are usually limited.
For example, a systematic review of the literature by Vidoni~\cite{vidoni_systematic_2022}  showed that approximately 37\% of the reviewed manuscripts do not address their repository selection process and do not detail the selection steps.
Moreover, Baltes and Ralph~\cite{baltes_sampling_2022} observed that discussions on sampling quality often rely on vague descriptors like “real-world”, “diverse,” or “representative,” but rarely provide supporting evidence. 
Developing tools to support the acquisition of representative samples, especially in complex scenarios involving \multistage sampling, may help address these challenges~\cite{maj_stars}. In their paper "The Fault in Our Stars"~\cite{maj_stars}, \citeauthor{maj_stars} emphasize the importance of explicitly motivating and justifying sampling strategies, demonstrating that different strategies can lead to substantially different conclusions.
Building on our earlier discussions, we believe that such tools should allow for the explicit specification of \multistage sampling processes, by offering enough flexibility to model diverse sampling strategies in the form of \textit{sampling workflows}. 
While one could manually depict such workflows using conventional drawing tools, we posit that enabling researchers to explicitly describe them by using well-defined operators provides both clarity and structural guidance, hence helping reduce the risk of error.
Guiding the explicit description of sampling workflows may indirectly support reproducibility, as extraction processes from software repositories are often insufficiently detailed in the literature~\cite{vidoni_systematic_2022}.
Moreover, an explicit representation %
enables automation and reasoning across the workflows' intermediate datasets and the operators connecting them. 
For instance, a grouping operator might compute and present how groups differ in metadata distributions relative to a sampling frame, while a random sampling operator could automatically estimate confidence intervals and margins of error based on sample size. 
Such capabilities would allow researchers to derive key indicators (e.g., sample size adequacy, parameter distributions) at each stage of the workflow, thereby supporting discussion for representativeness on well-documented and reproducible sampling strategies.

This paper aims to tackle the challenges described here by presenting a DSL for representation and reasoning on \multistage sampling strategies. Although the DSL we present can handle different type of \textit{Artifact}, this study focuses on \textit{Software Repository} artifact.

We further present a syntax and implementation enabling the sampling workflow to be actionable. Specifically, the sampling workflow can be executed on an initial dataset, while calculating appropriate indicators at each stage of the workflow to support the assessment of its representativeness. 
Our implementation relies on a Python fluent API to benefit from its ecosystem for statistics data analysis or graphical visualisation. 

Finally, we conducted a case study to evaluate the expressivity of our DSL on sampling strategy from the MSR literature. 
\section{A Domain Specific Language for modeling complex sampling workflow }

\begin{figure*}
    \centering
    \includegraphics[width=0.9\linewidth]{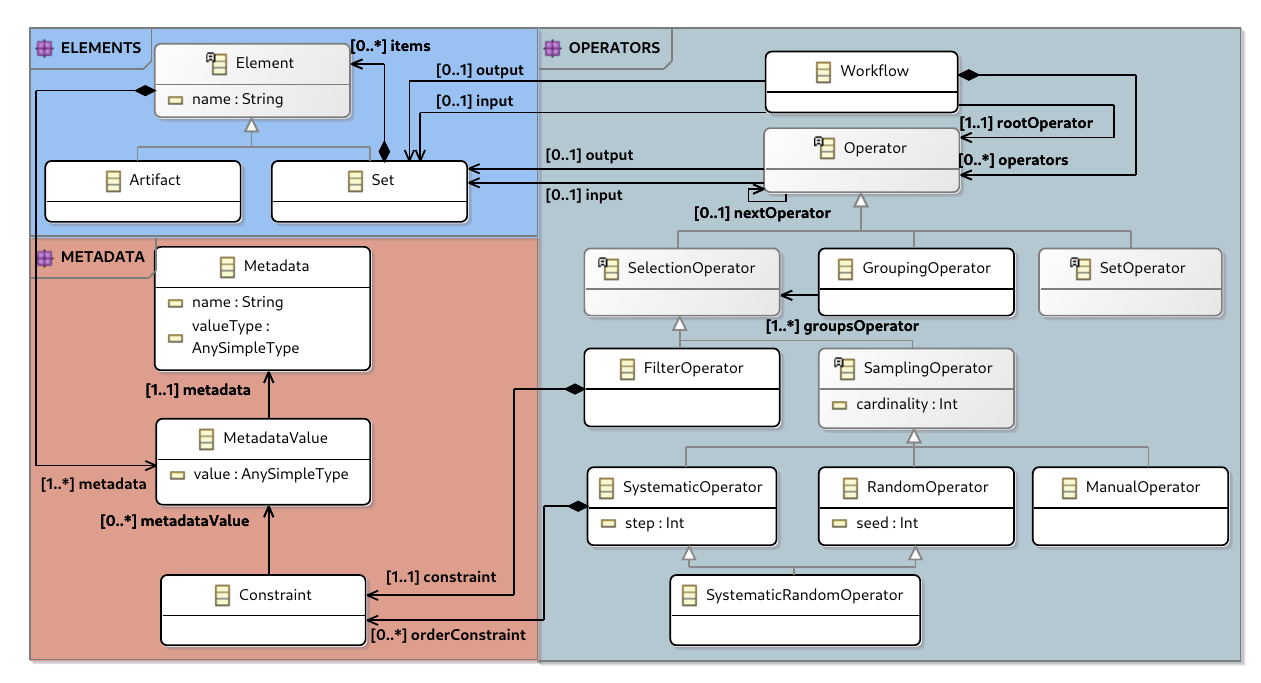}
    \caption{Metamodel of the Sampling Workflow Domain Specific Language}
    \Description{Metamodel of the Sampling Workflow Domain Specific Language}
    \label{fig:DSL}
\end{figure*}

\label{DSL}

To address the challenges discussed in the previous section, we propose a Domain Specific Language (DSL) to model sampling workflows. This DSL enables an explicit description of sampling workflows, such as \multistage sampling workflows that would not fit in the traditional three-tier framework. \Cref{fig:DSL} describes the metamodel, which specifies the abstract syntax (core concepts and their relationships) of our DSL. The metamodel is divided into three parts. The \textit{OPERATORS} of the workflow and the different relations between them. The \textit{ELEMENTS} on which the operators can be applied, and finally, the \textit{METADATA} associated to each \textit{Element}, as well as different constraints that could be expressed on them. The core contributions of this metamodel are the two first parts, the latter is intended to capture generic constraints and metadata store. In the remainder of this paper, we demonstrate the integration of software repository data sources.

\subsection{Concepts}

\paragraph{ELEMENTS}
\textit{Element} is an abstract concept we use to refer to an object  associated with different \textit{ metadata values} of a certain type of \textit{Metadata}. 
A software repository can be modeled as a subtype of \textit{Artifact} for which different values of software repository \textit{Metadata Value} can be associated, such as the language or the number of commits. A \textit{Set} is an \textit{Element} that contains a collection of distinct \textit{Elements}. 
A \textit{Set} can therefore be used to represent a dataset of repositories but also a partition of a given dataset.
\paragraph{OPERATORS}
A \textit{Sampling Workflow} is defined by the chaining of \textit{Operators}. An operator is a concept that describes different sampling strategies and operations, possibly chained, that could be performed on an input \textit{Set} and resulting in a modified output \textit{Set}. The rationale behind these concepts is to provide the basic \textit{Operator} to support the implementation of the sampling strategies of the literature (cf. \Cref{techniques}). For instance, complex strategies such as \textit{quota or clustering} sampling can be designed with the \textit{GroupingOperator} that can be chained to multiple \textit{Operators}.

Although generally chained to only one \textit{Operator}, some specific \textit{Operator} such  the \textit{Grouping Operator} can be chained to multiple \textit{Operators}.
A \textit{Selection Operator} supports the selection of \textit{Element} from the input set to the output set. Two concrete types of \textit{Selection Operator} are defined: \textit{Filter} and \textit{Sampling Operator}.
\textit{Filter Operator} refers to operator filtering \textit{Elements} of the input \textit{Set} through the execution of a \textit{Constraint} on a given \textit{Metadata Value}. The \textit{Filter Operator} is used to refine the sampling frame and reduce the scope of analysis, for instance, a \textit{ filter operator} might filter out all non-Java projects. \textit{Filter Operators} are generally used to scope and design sampling frame. The \textit{Sampling Operator} refers to operators  selecting a defined number of \textit{Elements}. \textit{Manual Sampling} is an operator that captures convenience sampling where elements are selected based on expediency or the expertise of the DSL user. The \textit{Random Operator} is the classical \textit{Random Sampling} operator and could be parameterized with a given seed.  The \textit{Systematic Operator} covers  systematic sampling, where elements are ordered with an \textit{Order Constraint} and selected at regular intervals.

\textit{Grouping Operator} is an \textit{Operator} composed of multiple \textit{Filter Operators} extracting different subsets of the input \textit{set}. A \textit{Grouping Operator} can be chained in different ways. In fact, each subset can be chained with any \textit{Operator}, the final result of the \textit{Grouping Operator} can also be chained. The execution of such an operator increases the depth of the output \textit{Set} (for instance, a \textit{Set} of \textit{Set} of \textit{Repositories}). Note that the number of subsets in the output \textit{Set} is equal to the number of \textit{Filter Operators}.

\textit{Set Operators} are operators for manipulating \textit{Set} composed of subsets, e.g., a union on the different subsets can be performed after a grouping operator.

Common multi-stage sampling approaches described in the literature (cf. \Cref{background_challenges}) can be implemented as composite operators:

The \textit{Cluster Operator}, corresponding to cluster sampling, can be implemented by chaining a \textit{Grouping Operator} with a \textit{Random Operator} to randomly select subsets of the grouping, which are considered as clusters.

The \textit{Stratified Random Operator}, corresponding to stratified sampling, can be implemented by chaining each subset of a \textit{Grouping Operator} with a \textit{Random Operator}. The \textit{Stratified Random Operator} imposes a constraint on the subsets extracted by the \textit{Grouping Operator}, requiring that they form a partition of the input set.

The \textit{Quota Operator}, corresponding to quota sampling, can be implemented in the same way as the \textit{Stratified Operator}. Unlike stratified sampling, elements in each subgroup are not randomly selected but chosen purposively.

\paragraph{METADATA}
\textit{Elements} of our model, such as \textit{Artifact} or \textit{Set} are associated with \textit{Metadata Values}, on which certain \textit{Operators} can impose \textit{Constraints}. 
These concepts are intentionally high-level and are intended to support repository-related metadata database, as well as other metadata formats like RDF \footnote{\url{https://www.w3.org/RDF/}}. The same applies for the \textit{Constraint} concept on which simple constraint specification such as boolean expression can be used but also more complex query language such as SPARQL\footnote{https://www.w3.org/TR/sparql11-query/}. The primary focus of this paper is to support the sampling workflow, independently of the data source and the constraint language.

\subsection{DSL Implementation}

Depending on the targeted user, a DSL can be implemented through different forms. For the evaluation purpose we implemented the DSL through a Python Fluent API describing an internal DSL. Indeed, the intended users of the DSL are MSR study designers, who are likely familiar with Python.  Internal DSLs \cite{fowler2010domain} are a category of DSL implemented through a host language, such as a General Purpose Language (GPL). Internal DSL leverages the host GPL concept contrary to external DSL, on which each concept, even the general one, should be re-implemented. For instance, we will reuse the Boolean algebra and the Python expression to implement the \textit{Constraint} concept. In addition, Python has an important ecosystem of data analysis libraries that facilitate the implementation of workflow execution analysis. Lastly, relying on an Internal DSL enables reuse of the host ecosystem such as the parser, interpreter, or compiler or the IDE support. 
The DSL is described by a fluent API, a concept introduced by Fowler\cite{fowler2010domain} to refer to API having properties making them suitable for internal DSL implementation and more particularly workflows internal DSL. The Python fluent API is only one concrete syntax derived from this metamodel, provided for usability. Alternative syntaxes, including an external DSL, could be directly derived from the metamodel, for example via Langium\footnote{https://langium.org/}.
We provide a reproduction package archived on Software Heritage that includes the DSL implementation details\footnote{ \swhref{swh:1:rev:24c12922d4f726d4686bec5cfef16a7914c18f81;origin=https://github.com/RomainLefeuvre/MSR_2026_reproduction_package}{swh:1:rev:24c12922d4f726d4686bec5cfef16a7914c18f81}}.

The prototype proposes two default loaders, a json loader and a csv loader, instantiating a \textit{Set} of \textit{Artifacts}. To illustrate and execute our running example on a real dataset we created a specific loader, that leverages the Software Heritage Graph \cite{pietri_swh} to execute the workflow on the entire archive or a subset of the archive. To
date, Software Heritage (SWH) is the platform archiving continuously the most OSS
repositories, from the different forge (cf \cref{rw_sf}). The SWH Loader leverages the \textit{Rust SWH graph API}  \footnote{\url{https://docs.softwareheritage.org/devel/swh-dataset/graph/}} to dynamically load the needed \textit{Metadata Values}. Similarly to \textit{Metadata}, other kinds of \textit{Constraint} systems or query languages might be added.
\Cref{fig:example}, models our running example of sampling workflow, expressed with the Python Internal DSL.   
Line 2 of \Cref{fig:example} declares the input of the workflow, which is a subgraph of 27,1M repositories randomly extracted from the Software Heritage Dataset \textit{2024-05-16-history-hosting}. Then, line 4 declares a \textit{FilterOperator} that filters out repositories whose last commit was before January 1, 2023. Repositories are partitioned by the number of committers using a \textit{GroupingOperator} (lines 5, 8, 12), resulting in two \textit{Sets}. Then, 10,000 repositories are randomly selected from each stratum using the \textit{RandomSelectionOperator} (lines 9 and 13). Finally, a union of the two extracted samples is performed.

\begin{figure}[htbp]
\centering
\begin{minted}[frame=single, fontsize=\footnotesize,xleftmargin=1.5em, linenos]{python}
(WorkflowBuilder() 
.input(SwhLoader("2024-05-16-history-hosting-subgraph",url,
swh_id, commit_count, commiter_count, latest_commit_date)) 
    .filter_operator("latest_commit_date > datetime(2023,1,1)") 
    .grouping_operator(
        # First stratum: projects with less than 5 committers
        (WorkflowBuilder()
        .filter_operator("commiter_count < 5")
        .random_selection_operator(10000)),
        # Second stratum: projects with 5 or more committers
        (WorkflowBuilder()
        .filter_operator("commiter_count >=5")
        .random_selection_operator(10000)))
    .union_operator() # Merge the two samples
    .execute())
\end{minted}
\caption{Modeled workflow of our running example}
\Description{Modeled workflow of our running example}
\label{fig:example}
\end{figure}

\section {DSL-Based Representativeness Reasoning \label{reasoning}}

Having a metamodel and an associated DSL enables the description of execution semantics to automatically evaluate a sampling workflow to extract a sample. It also describes analysis based on this formalism. Thus, we can describe statistical analysis supporting generalization reasoning of findings observed on a sample to a sampling frame.  
Such generalization is possible only if the sample is considered as representative of the targeted sampling frame on the dimension(s) of interest. In this section, we will explore the different arguments for representativeness (cf. Section \ref{rep_samp}), which can be supported by analyzing the execution of a workflow. 

For illustrative purposes, we executed our running example over a subset of the Software Heritage Graph, composed of \sizeInputDataset repositories randomly extracted. This dataset is considered as our initial \textit{Sampling Frame}. The intermediate \textit{Sets} obtained through the workflow execution is presented on \Cref{fig:flow}.

\subsection{Similar distribution analysis}
A report on the distribution of metadata of interest can be produced to compare the distribution of the input and output set of an \textit{Operator}. 
We propose two levels of analysis, descriptive statistics and statistical tests, to compare distributions. Descriptive analysis provides a high-level analysis by comparing metrics such as mean or dispersion. Different forms of descriptive analysis can be supported, such as histograms or box plots on metadata of interest. 

\begin{figure*}[t]
        \centering

            \includegraphics[width=1\linewidth]{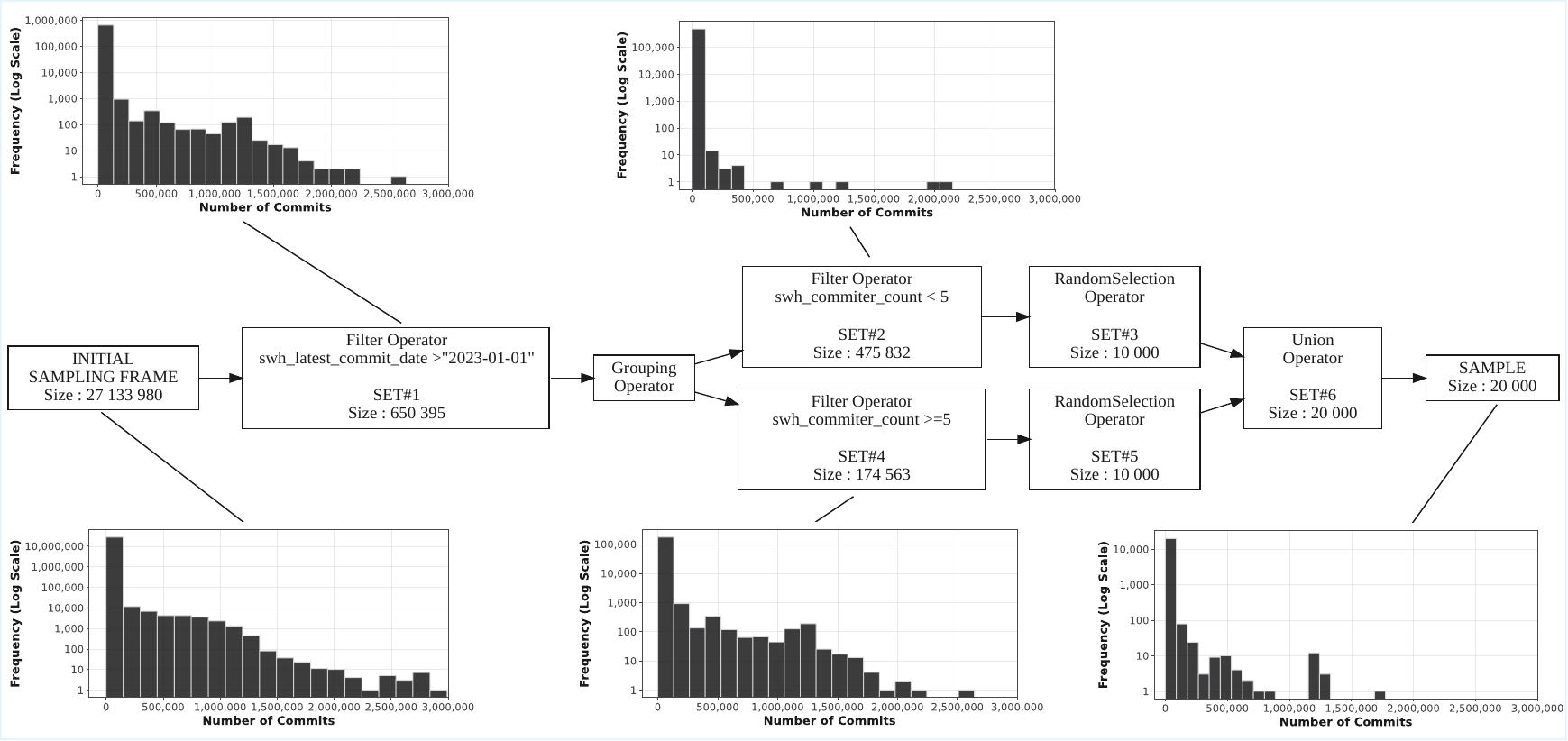}
\caption{Generated visualization of the workflow execution, with the distribution of the number of commits.}
\Description{Generated visualization of the workflow execution, with the distribution of the number of commits.}
        \label{fig:flow}
\end{figure*}

\Cref{fig:flow} presents the distribution of the number of commits, our metadata of interest, across the different intermediate \textit{Set} of the running example. To facilitate comparison between different sampling frames, the histograms are truncated to a maximum of 3 million commits. In the initial sampling frame, less than 0.001\% of commits are above this threshold. The histogram of the \textit{Initial Sampling Frame} and \textit{Set \#1}, filtered by \textit{latest commit date}, shows similar distributions, preserving the different classes of \textit{commit number}. This also applies when comparing the final sample to the \textit{Initial Sampling Frame}, indicating that balancing by \textit{committer number} using the grouping operator (\Cref{fig:example}, lines 5-13) maintains the sample's breadth on the metadata of interest.

In addition to visual analysis, statistical tests can be used to compare the distribution of a sample and a sampling frame depending on its nature. Among the statistical tests in the literature, the \textit{Chi-square goodness-of-fit} test, which allows the comparison of categories or enumeration metadata, can be used, for instance, for comparing distribution on metadata of interest such as repository main language. For continuous metadata such as commit number, the \textit{Kolmogorov-Smirnov} test, a non-parametric statistical test, can be used to compare the distributions of different sets, as in our example. Two pairs of sets have non-null p-values: 0.83 for (Set \#2, Set \#3) and 0.68 for (Set \#4, Set \#5). Both are superior to our significance level (0.05), the null hypothesis is not rejected, indicating that each pair of \textit{Set} does not have a different distribution.  This test strengthens the representativeness of each random sample with regard to the stratum they were extracted from.

\subsection{Statistical verification of sample size}

Different statistical approaches exist to estimate an acceptable sample size, depending on the sampling operator used and the distribution of the \textit{sampling frame} on metadata of interest, the size of the \textit{sampling frame} but also the type of analysis that will be conducted on the sample. 
We advocate that such a statistical test can be seen as an analysis, taking as input the execution of an operator, but also other parameters such as the metadata of interest or the type of analysis conducted, checking their applicability assumption, and returning whether or not the extracted sample is superior to the computed minimum sampling size. We will not present all kinds of statistical tests that could be applied on all the operators, nor discuss complex statistical tests, which are, in fact, a perspective of this study. We use \textit{Cochran’s formula }\cite{cochran1977sampling} to determine the minimum sample size needed, for a \textit{Random Operator} and its sampling frame, to achieve a desired confidence interval and ensure statistical reliability of the observations. Although this formula applies to an individual \textit{Random Operator}, variants using allocation techniques can be considered for composite operators, such as \textit{Stratified Random Operator} \cite{Singh1996}.

For our running example we calculated the minimal sample size for the two \textit{Random Operator} (\Cref{fig:example}, lines 9 and 13).
Such kind of analysis can be conducted thanks to the availability of the execution parameter of each \textit{Operator} in the workflow. Consequently, the sizes of the intermediate sets returned by the filter operators that precede each random operator (\Cref{fig:example}, lines 9 and 13) can be determined. In this case, they are 475,832 (Set \#2) and 174,563 (Set \#4). With a confidence level of 95\% and a margin of error of 0.05, the \textit{Cochran’s formula } yields 384 and 383, respectively. This suggests that the sample size of 10,000 is appropriate, thus supporting the representativeness of the produced random samples (Set \#3 and Set \#5) in relation to their respective stratum.

\section{Case Study}
\label{sec:casestudy}

In this section, we evaluate the applicability and expressiveness of our DSL by modeling sampling strategies from the MSR literature. 
We focus on this domain because MSR studies often rely on software forges, which, as discussed in Section 2, likely necessitate \multistage sampling strategies. 
Our goal is to assess whether the DSL can accurately capture these strategies and whether the resulting explicit representations can help identify ambiguities in their descriptions that may hinder reproducibility.
This study follows the guidelines for conducting and reporting case studies in SE~\cite{Runeson_Guideline} and is designed to answer the following research questions: 

\begin{itemize}
\item[\textbf{RQ1}] \textbf{What types of sampling strategies are used in the MSR literature? } We characterize the sampling approaches used and assess how frequently our solution applies.
    
\item[\textbf{RQ2}] \textbf{Do studies using sampling strategies include discussions on generalizability or representativeness?} We investigate the extent to which representativeness is addressed in the literature.
   
\item[\textbf{RQ3}] \textbf{Can the concepts defined in our DSL adequately capture the complexity of sampling strategies used in MSR studies?} We assess the expressiveness and completeness of our DSL. 

\item[\textbf{RQ4}] \textbf{What are the causes of incomplete modeling of sampling strategies?} 
We assess the causes of incomplete modeling that could be related to the proposed DSL concepts or the studied sampling strategy descriptions.

\end{itemize}

Note that this study does not focus on evaluating tool usability or effectiveness, as such assessments would be premature without first establishing the expressivity and conceptual validity of the DSL: user-based evaluations are left for future work.

\subsection{Dataset}

\begin{figure}[htbp]
\begin{subfigure}{0.46\textwidth}
\begin{minted}[frame=single, fontsize=\footnotesize,xleftmargin=1.5em, linenos]{python}
WorkflowBuilder()
    .input(CsvLoader(input_path,doi,title,year,numPages))\
    .filter_operator("2021 <= year <= 2025")\
    .filter_operator("numPages > 6")\
    .add_metadata(CsvLoader(IEEE_path,doi,ieee_keyword_list))\
    .random_selection_operator(cardinality=65,seed=4242)\                      
\end{minted}
\caption{Workflow source code }
\label{fig:case_study_workflow}
\end{subfigure}
\hfill
\begin{subfigure}{0.46\textwidth}
    \centering
    \includegraphics[width=\linewidth]{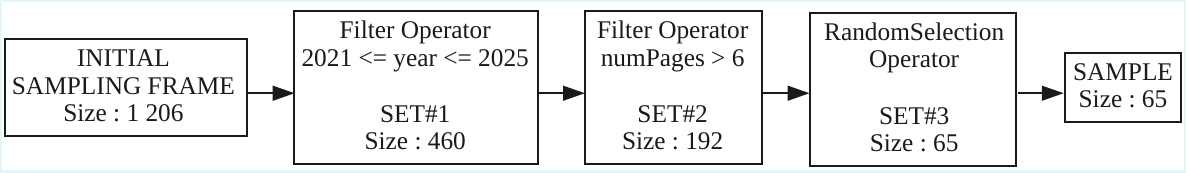}
    \caption{Generated Workflow diagram}
    \label{fig:case_study_workflow_diagram}
\end{subfigure}
\caption{Case study sampling workflow }
\Description{Case study sampling workflow }
\end{figure}

The ideal population of interest for this study consists of \textit{all empirical works that apply sampling to software repository artifacts}. As a sampling frame, we selected publications from the Mining Software Repositories (MSR) conference, whose domain makes it highly relevant for our study. 
To obtain a sample of these papers, we use our DSL (with scientific papers as \textit{Artifacts}) to describe our sampling strategy, presented in textual representation in \Cref{fig:case_study_workflow} and its associated generated diagram in \Cref{fig:case_study_workflow_diagram}. 
The initial sampling frame (\Cref{fig:case_study_workflow}, line 2) is composed of bibliographic information (in CSV format) extracted from dblp computer science bibliography \cite{dblp.xml.2025-07-02} SPARQL api\footnote{\url{https://sparql.dblp.org/}}.
The query used returned 1206 papers with the following metadata: \texttt{DOI}, \texttt{title}, \texttt{year} and \texttt{numPages}. 

The first operator in the workflow (line 3) filters papers from the last five years. We focus on this period to reflect the recent evolution of MSR research, including new development practices, expanded data sources, and advances in mining techniques and AI. Since our goal is to support current and future research, we restrict our analysis to this modern context, resulting in 460 papers. 
The next operator (line 4) filters long papers (over six pages), as they are more likely to include detailed and complete descriptions of sampling workflows. This helps avoid biased interpretations based on incomplete or underspecified methods due to lack of space in short papers. This step results in 192 papers. 
At this stage of the workflow (line 5), we load the IEEE keywords exported from the IEEE Xplore platform \footnote{https://ieeexplore.ieee.org/xpl/conhome/1001959/all-proceedings}. We use the IEEE classification keywords to check whether our sample adequately covers the MSR topics. 

The research questions described in the next section require manual analysis. Performing a full manual analysis of the entire data set would require substantial effort. To ensure feasibility while maintaining rigor, we applied random sampling with a confidence level of 95\% and a margin of error sufficient to support our interpretation. Using the \textit{Cochran's formula}, we determined that a sample of 64 papers would allow for this confidence level and margin of error. Accordingly, the last operator (line 6) randomly selects 65 papers, fulfilling the required sample size. A discussion of the representativeness of this sample is provided in the Threats to Validity section (Section \ref{sec:threat}).

\subsection{Methodology}

For each paper in the sample, we extracted the following data items (D1.1 to D4.2) to answer our research questions:

\textbf{RQ1: What types of sampling strategies are used in the MSR literature? 
}
 \textit{(D1.1) Presence of Code Repository Sampling Methodology}: Determine whether the study applies a sampling methodology to a software repository. This includes cases where the entire population is sampled (“whole frame”). 
 The remaining data items will be computed on the subset of papers conducting sampling on code repositories. 
 \textit{(D1.2) Data Sources}: Identify the software repository sources used in each study. This field lists all distinct data sources encountered.    \textit{(D1.3) Multistage Sampling:} Determine if the study utilizes a multistage sampling methodology. A sampling approach is defined as \multistage 
 if the associated modelled workflow is composed of multiple intermediate sampling frames. Workflows that use only \textit{filter operators} are classified as multistage only if they cannot be combined, e.g., when metadata are introduced in an intermediate sampling frame and require extensive computation and cannot be calculated on the initial sampling frame.

\textbf{RQ2: Do studies using sampling strategies include discussions on generalizability or representativeness? }
 \textit{(D2.1)/(D2.2) Discussion of Generalizability/Representativeness}: Determine whether the study discusses the generalizability of its findings (D2.1), and extract the relevant statements supporting this discussion (D2.2).

\textbf{RQ3: Can the concepts defined in our DSL adequately capture the complexity of sampling strategies used in MSR studies? }
 \textit{(D3.1) Sampling Process Description}: Extract all parts of the study that describe the sampling process applied to software repositories.
   \textit{(D3.2) Modeled Workflow}: Based on the extracted sampling process (D3.1), model the corresponding workflow using the DSL. If the process cannot be fully modeled, mark D4.1 as true. A modeling attempt is considered successful when all stages described in the original sampling process are captured by our DSL and when there is no ambiguity in operator behavior, metadata definitions, or constraints.
   
\textbf{RQ4: What are the causes of incomplete modeling of sampling strategies?
}
\textit{(D4.1) Incomplete Modeling}: Indicates whether a sampling workflow description has been fully modeled.
\textit{(D4.2) Cause of Incomplete Modeling}: Causes that prevent the complete modeling. Causes could be related to the expressivity of the DSL or the description of sampling strategies.

These data items were refined through an initial analysis of all MSR 2024 papers. Then, each paper in our sample was analysed by one author and independently reviewed by another author to reduce bias, particularly for data items requiring interpretation.

\subsection{Results }

\textbf{RQ1: What types of sampling strategies are used in the MSR literature?} Among the studied MSR papers, \DataOneOneYESPercent (\DataOneOneYESCount papers) perform sampling on code repositories, indicating that the majority of analysed sampling methodologies involve selecting code repositories.  
The remaining \DataOneOneNOPercent focus on other artifact types, such as code snippets, vulnerabilities, or nested repository artifacts, including releases, commits, files, or functions.  

\begin{figure}[htbp]
    \centering

        \includegraphics[width=0.9\linewidth]{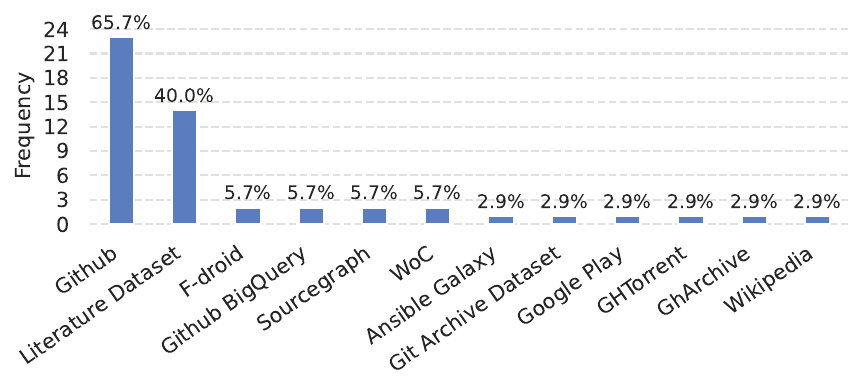}
        \caption{D1.2 Data sources - Frequency of data sources used in paper having code repository sampling. \small{(Percentage show the proportion of the \DataOneTwoTotal papers having such data source); a paper could use multiple data sources}}
        \Description{D1.2 Data sources - Frequency of data sources used in paper having code repository sampling. (Percentage show the proportion of the \DataOneTwoTotal papers having such data source); a paper could use multiple data sources}
        \label{fig:data-source}
    
\end{figure}

\Cref{fig:data-source} shows the different data sources used in the \DataOneOneYESCount papers that rely on code repositories. On average we observe  \DDatasourcesBoxplotMean data sources per paper, a median of \DDatasourcesBoxplotMedian and a maximum of 3. This indicates that most studies extract information from a single data source.  
Unsurprisingly, the results indicate that GitHub is the most commonly used data source, appearing in \DataOneTwoGithubPercent of the papers, close to the 67\% observed by \citeauthor{vidoni_systematic_2022}  \cite{vidoni_systematic_2022}. The second most used type of data source is literature datasets, employed by \DataOneTwoLiteratureDatasetPercent of the studies. This finding is particularly interesting, as it suggests that modeling sampling workflows and extracting representativeness arguments could also be valuable for other studies that reuse the dataset. 

The results of D1.3 reveal that \DataOneThreeYESPercent of the \DataOneThreeTotal papers that perform code repository sampling exhibit a multistage sampling methodology. This confirms that the sampling methodologies applied to code repositories are predominantly multistage in nature. This reflects the challenges discussed in Section~\ref{background_challenges} and reported in the literature~\cite{ma2018automatic,pfeiffer2020constitutes,vidoni_systematic_2022}.
The remaining \DataOneThreeNOPercent correspond to studies that do not employ multistage sampling. These include "whole frame" sampling strategy, where existing datasets from the literature are reused as-is, as well as workflows that are only composed of filter operators, without metadata computed on an intermediate sampling frame, which is semantically equivalent to a single \textit{Filter Operator}.

\textbf{RQ2: Do studies using sampling strategies include discussions on generalizability or representativeness?} The results of D2.1 show that \DataTwoOneYESCount papers out of \DataTwoOneTotal include a discussion on generalizability or representativeness. This shows that such considerations are common concerns among researchers. However, these discussions primarily focus on limitation of \textit{practical generalization},  addressing issues such as generalization to other programming languages, to other forges, or to closed-source projects. 
\begin{barquote}
\textit{
"Our analysis focused on Python repositories, excluding other languages that may have different characteristics. Additionally, our dataset consists of open-source GitHub projects"}
\end{barquote}

The literature-based arguments of representativeness described in \Cref{background_challenges} are rarely explicitly used with all the required assumption or statistical test. Our DSL can be leveraged to design sampling strategies for which statistical metrics supporting representativeness reasoning can be computed. 

For instance, studies that use as argument that their sample is \textit{"very large"} would leverage our DSL to automatically analyze \textit{Random Operators} and their associated sampling frames to compute statistical metrics such as \textit{Cochran’s formula}, ensuring that the sample size of \textit{Random Operators} is sufficient for a given level of confidence and margin of error.

Other studies discuss arguments that could be related to the \textit{breadth of a sample} or \textit{similar distribution} arguments and would benefit from statistical metrics. For example, the diversity of a project is used as an argument, such as "different programming languages and sizes" or "different frameworks and technologies." In these cases, our DSL enables the statistical comparison of the distribution or the coverage of language metadata and project size between the different sampling frames of the workflow.

We also observed cases where papers reusing datasets from the literature delegate the generalization reasoning to the original study, suggesting that generalization concerns can propagate to subsequent research. For example:

\begin{barquote}

\textit{"we selected the Git repositories for this study from the literature [..] and they already underwent a strict search and selection process, making us reasonably confident of their representativeness."}
\end{barquote}

In addition to the practical reasoning that authors use in analysed discussions, our DSL could support theoretical reasoning by providing statistical arguments to justify the representativeness of the extracted sample with regard to the available sampling frame.\\

\textbf{RQ3: Can the concepts defined in our DSL adequately capture the complexity of sampling strategies used in MSR studies? }Based on the sampling process descriptions (D3.1) extracted from each paper, we succeed to model \DataFourOneYESCount of the \DataOneOneYESCount  papers that described a sampling methodology for code repositories. It is worth to note that throughout this process, we did not identify any sampling operators that our DSL could not express, indicating that its base sampling operators are sufficient to represent the sampling methodologies used in our dataset. 

The remaining \DataFourOneNOCount papers that are not fully modeled, are not related to the DSL itself, but instead related to ambiguities in the descriptions of sampling strategies, which we discuss in RQ4.  

Thus, the concepts of our DSL enable us to capture the studied sampling strategies. All the concepts of the workflow are native to the language, capturing intrinsic complexity without introducing accidental complexity. The modeling of the sampling strategy enables us to provide a complete description of the workflow, contrary to textual representations that could be ambiguous.

\textbf{RQ4: What are the causes of incomplete modeling of sampling strategies? }
\begin{figure}
    \centering
    \includegraphics[width=0.9\linewidth]{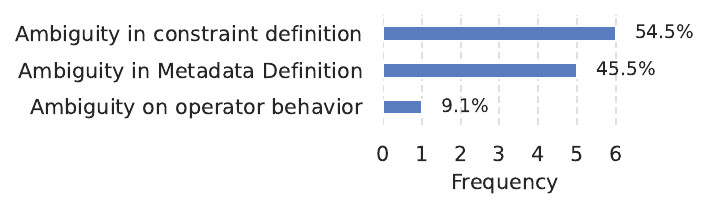}
    \caption{D4.2 Distribution of Causes \small{(The percentages correspond to the proportion of the \DataFourTwoTotal papers with a specific ambiguity cause; a paper can have multiple ambiguity causes).}}
    \Description{D4.2 Distribution of Causes (The percentages correspond to the proportion of the \DataFourTwoTotal papers with a specific ambiguity cause; a paper can have multiple ambiguity causes)}
    \label{fig:d42}
\end{figure}
Different ambiguities emerged during the modeling of \DataFourOneNOCount papers out of \DataFourOneTotal. 
\Cref{fig:d42} presents the distribution of the different causes of ambiguity. After analysis, 3 different causes of ambiguity have been identified. The main cause, representing \DataFourTwoAmbiguityinconstraintdefinitionPercent of cases, arises from missing information regarding the constraints used in the filter operator. Such ambiguity are caused by the usage of imprecise terms, for example \textit{a sufficient number of commits} where sufficient is not defined.

Another source of ambiguity, appearing in \DataFourTwoAmbiguityinMetadataDefinitionPercent of the cases, concerns metadata definition. We consider a metadata definition ambiguous when the described metadata does not clearly correspond to repository-level data (e.g., number of commits, number of committers) and the process used to obtain it is not explained in the paper or its references.

\begin{barquote}
\textit{"1. Accessibility, i.e., the libraries \textbf{should be popular},
widely used, and open-source, 2. Maturity, i.e., the libraries
should have been \textbf{actively developed} for a considerable amount
of time"}
\end{barquote}

In this citation, we observe that several constraints rely on metadata that are not explicitly defined. For example, popularity could be measured using GitHub stars or forks, while "developed for a considerable amount of time" might refer to the commit activity of a repository, which can be operationalized in various ways \cite{10.1145/2597073.2597074}. 

The last cause of ambiguity we identified concerns the operator's behavior. In the following excerpt, the workflow describes a process intended to balance the distribution of two samples:
\begin{barquote}
\textit{[..] we mitigate the effects of potential confounds in our analysis, for example, \textbf{choosing a similar amount of small, medium, and large-sized projects for both ML and non-ML projects}.}
\end{barquote}

The description specifies that project size should be balanced across samples but does not clarify how items are selected (e.g., randomly or purposively). This could be implemented using various strategies, such as stratified random or quota sampling.

In previous cases, the lack of explicit description prevents unambiguous interpretation and implementation of the sampling process.

\subsection{Threats to validity}
\label{sec:threat}

We could not discuss the representativeness of MSR studies without reasoning about the representativeness of our own sample.
As described in the preceding section, we focused on long papers published in MSR between 2021 and 2025. Thanks to the workflow modeling, we automatically computed several statistical tests that could be reused for representativeness analysis. We applied a sample size analysis (using \textit{Cochran’s formula}) to determine a size that ensures an acceptable margin of error and confidence level while remaining manageable for analysis. Exploratory studies generally adopt a 10\% margin of error and a 95\% confidence level. The result suggested a sample size of 64, and we selected 65 papers accordingly. Statistically, this means that the findings derived from our four research questions can be expected to represent a larger population of relevant articles with a reasonable level of confidence for an exploratory study—that is, if we analyze different samples of the same size, 95\% of the time the results will fall within 10\% of the population value. Such a margin would not change the interpretation of our research questions. These interpretations are possible under the assumption that the confidence level and margin of error remain consistent across the different research questions, since they are all based on the same sample. However, reusing this sample across multiple research questions, and considering research questions that focus on subsets of the sample, may affect their individual confidence levels and margins of error.

We also examined whether our sample is representative of long MSR papers published between 2021 and 2025 by analyzing the topics discussed in each. As sampling practices may differ across topics, we added the IEEE-provided keywords for each paper before performing random sampling (see line 5 of \Cref{fig:case_study_workflow}). We conducted two analyses. First, we verified the coverage of the 50 most frequently used keywords in long MSR papers between 2021 and 2025 and observed 100\% coverage in our final sample. Second, we compared keyword distributions before and after random sampling. The \textit{Chi-square goodness-of-fit} test returned a chi-square value of 197.3 and a p-value of 0.99, indicating no significant difference. Thus, our sample appears representative in terms of study topics. This discussion on generalization and representativeness defines the external validity of our study. Our findings cannot be generalized beyond long MSR papers published between 2021 and 2025; they may not apply to short papers or earlier studies.

To mitigate construct validity threats and potential bias in data item selection, we followed established guidelines for conducting case studies in software engineering \cite{Runeson_Guideline}. Different data items, \textit{such as D3.2 (Workflow Modeling)} or \textit{D4.2 (Cause of Incomplete Modeling)}, involved interpretation by the authors.

Internal validity could be affected by researcher bias, since the same team that designed the DSL also performed the workflow modeling and analysis. To reduce this risk, all extractions and interpretations were cross-checked by a second author, and disagreements were discussed until consensus was reached. We also verified the consistency of the DSL encoding through iterative review of the models. Automation errors in parsing or metadata extraction were minimized by manually validating intermediate outputs.

Finally, regarding conclusion validity, our results rely primarily on qualitative assessments and descriptive statistics. Although this is appropriate for an exploratory study, we acknowledge that the interpretation of “expressiveness” or “representativeness reasoning” could vary among evaluators. To enhance transparency and replicability, we provide a complete replication package that allows re-execution of the workflow, reproduction of all statistical analyses, and independent verification of the reported findings. Future work will include user studies to further validate the DSL’s usability and its impact on reasoning about representativeness.

\section{Related Work \label{related_work}}
\paragraph{Related work related to sampling in SE}
\label{rw_sf}
\citeauthor{maj_stars} question a widespread practice in MSR: the use of stars as a sampling criterion \cite{maj_stars}. Through a case study, they demonstrate that samples composed of starred projects are not necessarily representative of the target population. They argue for the design of sampling methodologies based on intrinsic metadata, rather than extrinsic metadata such as stars. In addition, they propose a conceptual framework consisting of multiple stages, including a description of the sampling strategy and a discussion of the representativeness of the selected sample. Our work complements their framework. The DSL we propose not only allows for modeling the sampling strategy, but also enables automatic analysis of the workflow, producing statistical metrics to support discussions about sample representativeness.

\citeauthor{Jordi_sample} propose an approach based on stratified random sampling
in which the strata are computed using a clustering algorithm such as k-means 
\cite{Jordi_sample}. Their approach could be integrated into our DSL by creating a new operator. 
Similarly,
~\citeauthor{Nagappan_2013_diversity}  propose a technique to assess the coverage of a sample of software projects in multiple~\cite{Nagappan_2013_diversity}.

Other work, such as \cite{carruthers2024longitudinal}, investigates the temporal validity of a sample by studying how the distributions of properties of interest evolve over time. After analyzing different properties of a sample of 1,991 repositories, they performed a survival analysis and found that, for all properties of interest, the probability that the distribution remains unchanged drops below 25\% after five years. In the context of maintaining representative samples and datasets over time, their study highlights the importance of being able to re-execute sampling strategies to extract updated samples, as well as to recompute statistical indicators to support representativeness reasoning.
Our DSL facilitates the re-execution of sampling strategies by making the sampling workflow explicit and by providing automatic statistical indicators, such as distribution analyses.

\paragraph{Related work related to SE sampling frame}
Beyond the sampling methodology, the initial sampling frame in which the sampling workflow is applied is critical for generalization reasoning.  Various contributions in the literature focus on providing broader sampling frames, but also the tooling to properly query or compute/extract metadata from this sampling frame. Using a broader or multiple sampling frames that does not focus only on a subset of the population is one of the practical arguments that can be used when it comes to reason on the representativeness of the targeted group to the population of interest; cf. \Cref{fig:generalization}. Indeed, many studies focusing on repository in SE are based on Github, however, the representativeness of GitHub to OSS population can be discussed.  
Software Heritage (SWH)\cite{pietri_swh} collects and preserves OSS with the aim of building a universal archive of source code along with its development history.
OSS projects are collected from public forges, with the objective to crawl as many forges as possible, including smaller forges hosted by various organizations. To date, SWH is the platform that is continuously archiving the most OSS repositories from different forges. It also offers the possibility to deploy locally an infrastructure to query the graph, through a unified low-level API (SWH-GRAPH). Therefore, using SWH as initial sampling frame is equivalent to querying multiple sampling frames (i.e., different forges) simultaneously, enhancing the representativeness of this sampling frame for populations such as OSS.

Other works also aim to provide the ability to query broader sampling frames or to overcome limitations associated when it comes to query commonly used sampling frames in SE, such as GitHub.  World Of Code~\cite{WOC} is an infrastructure that is archiving OSS with the objective to make it queryable by researchers. It covers various forges such as GitHub, Gitlab and Bitbucket, and allows users to build queries based on maps of different elements within the software graph. It is currently one of the most widely used sampling frames in the MSR community, offering a high-level API. %

The expressivity of the queries that can be performed on sampling frame can directly impact on the chosen sampling workflow and, consequently, the representativeness of a sample to its sampling frame. For instance, applying random sampling on a sampling frame requires being able to easily enumerate all the elements of the sampling frame, which can be challenging with GitHub. 

Beyond the ability to query repositories or any fine-grained artifacts, constructing metadata during the traversal allows the design of complex metadata of interest can be leveraged for sampling. Although it is possible to construct such metadata with SWH, it still relies on the use of low-level APIs. Boa~\cite{BOA} proposes a DSL to define analysis tasks in the context of mining software repositories. It allows querying, but also to construct, directly in the DSL, metadata while traversing repository graph structure. They provide an infrastructure to execute queries on different sampling frames, though it does not match the scale of SWH or World of Code.

\paragraph{Related work related to scientific workflow systems}

Scientific Workflows Systems (SWSs) enable the modeling and orchestration of heterogeneous  computational tasks to achieve a research, from data preprocessing to analysis \cite{alam2025empirical}. SWSs are key to handle reproducibility and replicability challenges of scientific experiments \cite{munafo2017manifesto,Felix_2021_sustainable}. 
SWSs can be categorized into different “niches” according to the level of abstraction offered to researchers \cite{Felix_2021_sustainable}. Some SWSs provide accessible graphical interfaces, such as KNIME \cite{knime_2008} or Galaxy \cite{galaxy2022galaxy}, which can be used by non-programmer researchers. Other SWSs are provided either as programming-language frameworks, such as Anduril \cite{cervera2019anduril}, offering extensibility and modularity, or as domain-specific languages (DSLs), such as Snakemake \cite{koster2012snakemake} or Nextflow \cite{di2017nextflow}. 
SWSs model and support the execution of the entire experimentation process, whereas our approach focuses on the design and justification of sampling strategies. In this sense, sampling workflows can be regarded as a subtype of scientific workflows. Our DSL supports the direct computation of indicators essential for arguing representativeness, which are not natively available in SWSs like KNIME\cite{knime_2008}. For instance, the proposed DSL automates the calculation of \textit{Cochran's formula} to verify sample size adequacy at specific stages and performs \textit{Kolmogorov-Smirnov} test for distribution similarity between intermediate sets.
\section{Conclusion and Perspectives }
\label{conclusion}
In this work, we introduced a metamodel to systematically represent sampling workflows of code repositories, addressing a key need for reproducibility and transparency in handling large datasets. This metamodel allows researchers to construct detailed, \multistage sampling workflows, supporting the rigorous design of samples that aim to accurately represent broader populations. We provided a Domain-Specific Language (DSL) to implement this metamodel, demonstrating its expressiveness through a comprehensive case study analyzing \DataOneOneTotal MSR papers with \DataOneOneYESCount that exhibit code repository sampling. Furthermore, we explored how statistical analyses of these workflows could support reasoning about the generalizability of findings, helping researchers better argue the validity of inferences drawn from sample data.

Several promising directions for future work build upon this foundation. 
Enhancing the metamodel to include fine-grained artifacts, such as commits and individual changes, would increase its adaptability for studies that require deeper levels of data granularity.
To further broaden its applicability, the prototype could be optimized for large-scale datasets and integrated with diverse metadata sources. For example, the integration of Software Heritage (SWH) graph API could be enhanced to integrate all the metadata available, improving the flexibility of the framework in handling diverse sampling needs. Finally, future work could explore the integration of our DSL with general SWSs.
 
\begin{acks}
This research was partially supported by the Brittany Region, France. %
\end{acks}
\newpage
\bibliographystyle{ACM-Reference-Format}
\bibliography{bib}
\end{document}